\documentclass[11pt,draftcls, onecolumn]{IEEEtran}

\usepackage{amsmath} \allowdisplaybreaks
 \usepackage{changebar}
\usepackage{amssymb}

\ifCLASSINFOpdf
      \usepackage[pdftex]{graphicx}
\else
\usepackage[dvips]{graphicx}
\fi

\ifCLASSOPTIONcompsoc
\usepackage[caption=false,font=normalsize,labelfont=sf,textfont=sf]{subfig}
\else
\usepackage[caption=false,font=footnotesize]{subfig}
\fi

\usepackage{array}
\usepackage{multirow}  % multiple row in table
\usepackage{array} % fixed talbe
\newcommand{\PreserveBackslash}[1]{\let\temp=\\#1\let\\=\temp}
\newcolumntype{C}[1]{>{\PreserveBackslash\centering}p{#1}}
\newcolumntype{R}[1]{>{\PreserveBackslash\raggedleft}p{#1}}
\newcolumntype{L}[1]{>{\PreserveBackslash\raggedright}p{#1}}
\usepackage{chemarrow}
\usepackage{mdwmath}
\usepackage{stfloats}

\usepackage[lined,boxed,commentsnumbered, ruled]{algorithm2e}

\begin{document}
\title{Improved \emph{N}-continuous OFDM for 5G Wireless Communications}

%%% title: 标题
%%%   \title{full title}{title for citation}
\author{Peng~Wei, Lilin~Dan, Yue~Xiao, Wei~Xiang, and Shaoqian~Li% <-this % stops a space
%%\thanks{This paper was presented in part at the IEEE International Conference on Communications, Budapest, CA, Hungary, Jun. 2013.}
\thanks{The authors are with the school of National Key Laboratory of Science and Technology on Communications, University of Electronic Science and Technology of China, Chengdu, China (e-mail: wpwwwhttp@163.com; \{lilindan, xiaoyue\}@uestc.edu.cn). Wei Xiang is with the school of Mechanical and Electrical Engineering Faculty of Health, Engineering and Sciences, University of Southern Queensland, Austrialia (e-mail: wei.xiang@usq.edu.au).}}%

%%% Abstract: 摘要
%%% Not necessery for LETTER. LETTER不需要摘要
\maketitle

\begin{abstract}
\emph{N}-continuous orthogonal frequency division multiplexing (NC-OFDM) is a promising technique to obtain significant sidelobe suppression for baseband OFDM signals, in future 5G wireless communications. However, the precoder of NC-OFDM usually causes severe interference and high complexity. To reduce the interference and complexity, this paper proposes an improved time-domain \emph{N}-continuous OFDM (TD-NC-OFDM) by shortening the smooth signal, which is linearly combined by rectangularly pulsed OFDM basis signals truncated by a smooth window. Furthermore, we obtain an asymptotic spectrum analysis of the TD-NC-OFDM signals by a closed-form expression, calculate its low complexity in OFDM transceiver, and derive a closed-form expression of the received signal-to-interference-plus-noise ratio (SINR). Simulation results show that the proposed low-interference TD-NC-OFDM can achieve similar suppression performance but introduce negligible bit error rate (BER) degradation and much lower computational complexity, compared to conventional NC-OFDM.
\end{abstract}

%%% Keywords: 关键词
%%% Not necessery for LETTER. LETTER不需要关键词

\IEEEpeerreviewmaketitle

%%%%%%%%%%%%%%%%%%%%%%%%%%%%%%%%%%%%%%%%%%%%%%%%%%%%%%%
%%% The main text. 正文部分
%%% No section name for LETTER. LETTER不分章节
%%%%%%%%%%%%%%%%%%%%%%%%%%%%%%%%%%%%%%%%%%%%%%%%%%%%%%%
%=======================================================================================================================================================================
\section{Introduction}

Orthogonal frequency division multiplexing (OFDM) \cite{Ref1} has been one of the most popular multicarrier transmission techniques in future 5G wireless communications \cite{Ref18, Ref19} due to its high-speed data transmission and inherent robustness against the inter-symbol interference (ISI). However, in rectangularly pulsed OFDM systems, the signal possesses a discontinuous pulse edge and thus exhibits large spectral sidelobes. Thus, power leakage due to sidelobes, which is also known as out-of-band power emission, causes severe interference to adjacent channels \cite{Ref2}, especially in cognitive radio (CR) and carrier aggregation (CA) combined 5G systems \cite{Ref20, Ref21}.

For improving conventional OFDM in out-of-band emission, various methods have been proposed for sidelobe suppression [7--18]. %\cite{Ref3,--,Ref14}
The windowing technique in \cite{Ref3} extends the guard interval in the price of a reduction in spectral efficiency. Cancellation carriers \cite{Ref4, Ref5} consume extra power and incur a signal-to-noise ratio (SNR) loss with high complexity. In the precoding methods [10--12], %\cite{Ref6-Ref8}, 
complicate decoding algorithms are required to eliminate the interference caused by the precoders. 

NC-OFDM techniques [13--19] %\cite{Ref9-Ref15} 
smooth the amplitudes and phases of the OFDM signal by making the OFDM signal and its first \emph{N} derivatives continuous (so-called \emph{N}-continuous). Conventional NC-OFDM \cite{Ref9} obtains the \emph{N}-continuous signal at the expense of high interference. Aiming at optimizing the frequency domain precoder in \cite{Ref9}, Beek et al. proposed the memoryless scheme in \cite{Ref10} and the improved scheme \cite{Ref11} nulling the spectrum at several chosen frequencies. On the other hand, to enable low-complexity signal recovery in NC-OFDM, several approaches have been proposed [16--18]. %\cite{Ref12-Ref14}. 
However, similar to the precoding techniques, the existing NC-OFDM techniques need robust signal recovery algorithms for reception. Among them, some methods degrade system performance, such as peak-to-average-power ratio (PAPR) growth in \cite{Ref10} and high complexity of transmitter in \cite{Ref12, Ref13}.

In this paper, to reduce the interference of the NC-OFDM signals and obtain a low-complexity transceiver, we propose a low-interference time-domain \emph{N}-continuous OFDM (TD-NC-OFDM) based on the conventional TD-NC-OFDM \cite{Ref15}. A smooth signal is superposed in the front part of each OFDM symbol to achieve \emph{N}-continuous OFDM signal, including smoothing both edges of the transmitted signal. The smooth signal is linearly combined by the basis vectors in a basis set, which is composed of the rectangularly pulsed OFDM basis signals truncated by a smooth window function for an example. Furthermore, we give rise to analyses of spectrum, complexity and signal-to-interference-plus-noise ratio (SINR) in the low-interference TD-NC-OFDM. Among them, an asymptotic expression of PSD of the TD-NC-OFDM signal is first obtained, where the sidelobes asymptotically decay with $f^{-N-2}$, when the first \emph{N} derivatives of the OFDM signal are all continuous. Then, we compare the complexity among NC-OFDM, TD-NC-OFDM and its low-interference scheme, to show the complexity reduction of the proposed low-interference scheme. The closed-form expression of the received SINR of the low-interference scheme is also calculated to show the slight SNR loss. Simulation results show that the low-interference scheme can achieve as notable sidelobe suppression as NC-OFDM method \cite{Ref9} with excellent bit error rate (BER) performance and low complexity.  

The remainder of the paper is organized as follows. In Section 2, the OFDM signaling is briefly introduced, the traditional NC-OFDM is reviewed. Section 3 proposes the low-interference TD-NC-OFDM model, gives the linear combination design of the smooth signal with a new basis set, and describes the transmitter. In Section 4, the effect of the low-interference TD-NC-OFDM on sidelobe decaying and the received SINR is analyzed as well as the computational complexity of the transceiver. Finally, Section 5 draws concluding remarks.

\emph{Notation}: Boldfaced lowercase and uppercase letters represent column vectors and matrices, respectively. $\{\mathbf{A}\}_{m,n}$ indicates the element in the \emph{m}th row and \emph{n}th column of matrix \textbf{A}. The $M\times M$ identity matrix and $M\times N$ zero matrix are denoted by $\mathbf{I}_M$ and $\mathbf{0}_{M\times N}$, respectively. $|\cdot|$ represents the absolute value. The trace and expectation of a matrix are represented by $\mathrm{Tr}\{\cdot\}$ and $E\{\cdot\}$, respectively. $\mathbf{A}^T$, $\mathbf{A}^{\ast}$, $\mathbf{A}^H$ and $\mathbf{A}^{-1}$ denote the transposition, conjugate, conjugate transposition, and inverse of matrix \textbf{A}, respectively. 

%=======================================================================================================================================================================
\section{System aspects and \emph{N}-continuous OFDM}

\subsection{OFDM signaling}
In a baseband OFDM system, the input bit stream of the \emph{i}th OFDM symbol is first modulated onto an uncorrelated complex-valued data vector $\mathbf{x}_i={[ x_{i,k_0}, x_{i,k_1},\ldots,x_{i,k_{K-1}}]}^T$ drawn from a constellation, such as phase-shift keying (PSK) or quadrature amplitude modulation (QAM). The complex-valued data vector is mapped onto \emph{K} subcarriers with the index set $\mathcal{K}=\left\{k_0,k_1,\ldots,k_{K-1}\right\}$. An OFDM signal is formed by summing all the \emph{K}-modulated orthogonal subcarriers with equal frequency spacing $\Delta f=1/T_s$, where $T_s$ is the OFDM symbol duration. The \emph{i}th OFDM time-domain symbol, assuming a normalized rectangular time-domain window $R(t)$ \cite{Ref6}, can be expressed as
\begin{equation}
  y_i(t)=\sum\limits^{K-1}_{r=0}{x_{i,k_r}e^{j2\pi k_r\Delta ft}}, -T_{cp}\leq t <T_s
  \label{Eqn:1}
\end{equation}
where $T_{cp}$ is the cyclic prefix (CP) duration. Then, in the time range of $(-\infty,+\infty)$, the transmitted OFDM signal $s(t)$ can be written as
\begin{equation}
  s(t)=\sum\limits^{+\infty}_{i=-\infty}{y_i\left(t-iT\right)}.
  \label{Eqn:2}
\end{equation}
where $T=T_s+T_{cp}$. 

After the OFDM signal is oversampled by a time-domain sampling interval $T_{samp}=T_s/M$, the discrete-time OFDM signal is expressed as
\begin{equation}
y_i(m)=\frac{1}{M}\sum\limits^{K-1}_{r=0}{x_{i,k_r}e^{j2\pi \frac{k_r}{M}m}},
  \label{Eqn:3}
\end{equation}
where $m\in \mathcal{M}=\left\{-M_{cp},\ldots,0,\ldots,M-1\right\}$, and $M_{cp}$ is the length of CP samples.

\subsection{\emph{N}-continuous OFDM}

To improve the continuity of the time-domain OFDM signal, the conventional NC-OFDM \cite{Ref9} introduces a frequency-domain precoder to making the OFDM signal and its first \emph{N} derivatives continuous. NC-OFDM follows that
\begin{equation}
  \bar{y}_i(t)=\sum\limits^{K-1}_{r=0}{\bar{x}_{i,k_r}e^{j2\pi k_r\Delta ft}}, -T_{cp}\leq t <T_s
  \label{Eqn:4}
\end{equation}
\begin{equation}
  \left. \bar{y}^{(n)}_i(t)\right|_{t=-T_{cp}}=\left. \bar{y}^{(n)}_{i-1}(t)\right|_{t=T_s},
  \label{Eqn:5}
\end{equation}
where $\bar{x}_{i,k_r}$ is the precoded symbol on the \emph{r}th subcarrier, and $\bar{y}^{(n)}_{i}(t)$ is the \emph{n}th-order derivative of $\bar{y}_{i}(t)$ with $n\in\mathcal{U}_N=\{0,1,\ldots,N\}$. 

\begin{figure}[h]
\centering
\includegraphics[width=3in]{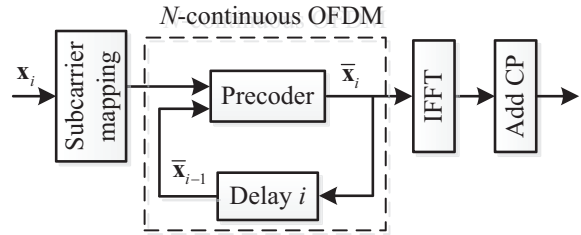}
\DeclareGraphicsExtensions.
\caption{Block diagram of the \emph{N}-continuous OFDM transmitter.}
\label{Fig:1}
\end{figure}

Based on \eqref{Eqn:4} and \eqref{Eqn:5}, the precoding process can be summarized as
\begin{equation}
\begin{cases}
 \bar{\mathbf{x}}_i=\mathbf{x}_0, & i=0  \\
 \bar{\mathbf{x}}_i=(\mathbf{I}_K-\mathbf{P})\mathbf{x}_i+\mathbf{P}\mathbf{\Phi}^H\bar{\mathbf{x}}_{i-1}, & i>0
\end{cases}
  \label{Eqn:6}
\end{equation}
where $\mathbf{I}_K$ is the identity matrix, $\mathbf{P}=\mathbf{\Phi}^H\mathbf{A}^H(\mathbf{A}\mathbf{A}^H)^{-1}\mathbf{A}\mathbf{\Phi}$, $\left\{\mathbf{A}\right\}_{n+1,r+1}=k^n_r$, $\mathbf{\Phi}=diag(e^{j\varphi k_0},e^{j\varphi k_1},\ldots,e^{j\varphi k_{K-1}})$, and  $\varphi=-2\pi\beta$ with $\beta=T_{cp}/T_s$. Figure \ref{Fig:1} depicts the spectrally precoded NC-OFDM transmitter. The \emph{i}th frequency-domain data vector $\mathbf{x}_i$ is first precoded. The precoded data vector $\bar{\mathbf{x}}_i=\left[\bar{x}_{i,k_0},\bar{x}_{i,k_1},\ldots,\bar{x}_{i,k_{K-1}}\right]^T$ then undergoes the inverse fast Fourier transform (IFFT), and finally the CP is added to generate the transmission signal.

%=======================================================================================================================================================================
\section{Proposed Low-Interference Scheme of TD-NC-OFDM}

\begin{figure}[htbp]%[!t]
\centering
\includegraphics[width=3.5in]{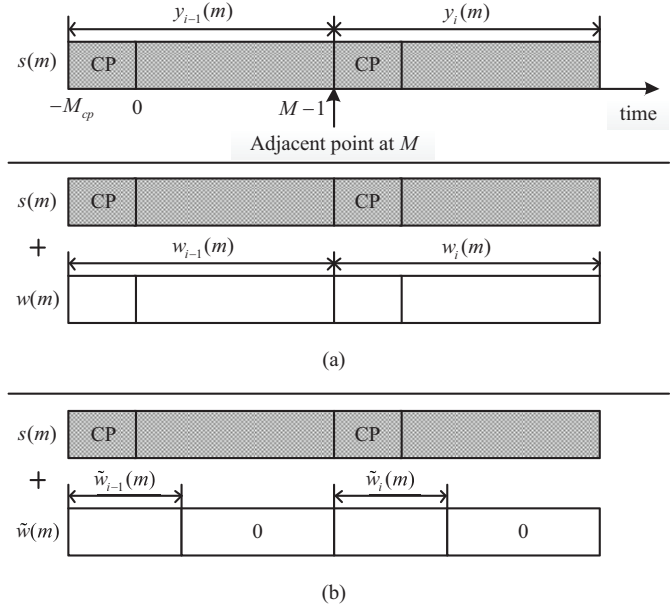}
\DeclareGraphicsExtensions.
\caption{Conventional and proposed ways of of adding the smooth signal in the time domain: (a) Addition in the whole OFDM symbol; (b) Addition in the front of each OFDM symbol.}
\label{Fig:2}
\end{figure}

Conventional TD-NC-OFDM \cite{Ref15} and Conventional NC-OFDM \cite{Ref9} cause high interference, which is needed to be reduced by complicate signal recovery algorithms [13, 16, 18]. %\cite{Ref9, Ref12, Ref14}. 
As show in Figure \ref{Fig:2} (a), the interference term, defined as the smooth signal $w_i(m)$, is located in the whole time-domain in the conventional TD-NC-OFDM as well as NC-OFDM . To eliminate the interference and simplify the receiver, we truncate $w_i(m)$ with a window function. As illustrated in in Figure \ref{Fig:2} (b), the truncated term $\tilde{w}_i(m)$ only locates in the front section of each OFDM symbol. 

To make the OFDM signal \emph{N}-continuous, $\tilde{w}_i(m)$ should satisfy 
\begin{equation}
  \bar{y}_i(m)=y_i(m)+\tilde{w}_i(m),
  \label{Eqn:7}
\end{equation}
\begin{equation}
  \left.\tilde{w}^{(n)}_i\!(m)\right|_{m=-M_{cp}}\!\!=\left.y^{(n)}_{i-1}\!(m)\right|_{m=M}
  \!-\left.y^{(n)}_{i}\!(m)\right|_{m=-M_{cp}}\!.
  \label{Eqn:8}
\end{equation}

$\tilde{w}_i(m)$ is the linear combination of the first \emph{N}+1 basis vectors in a basis set $\mathcal{Q}$, written as
\begin{equation}
  \tilde{w}_i(m)=\left\{\begin{matrix}
          \sum\limits^{N}_{n=0}{{b}_{i,n}\tilde{f}_n(m)}, & m\in \mathcal{L} \\
          0, & m\in \mathcal{M}-\mathcal{L}
  \label{Eqn:9}
\end{matrix}\right.,
\end{equation}
where $\mathcal{L}=\left\{-M_{cp},-M_{cp}+1,\ldots,-M_{cp}+L-1\right\}$ indicates the location of $\tilde{w}_i(m)$ with the length of \emph{L}, and the basis set $\mathcal{Q}$ is given by 
\begin{equation}
%\begin{aligned}
  {\mathcal{Q}}=\left\{{\mathbf{q}}_{\tilde{n}}\left|{\mathbf{q}}_{\tilde{n}}=\left[\tilde{f}_{\tilde{n}}(-M_{cp}),\tilde{f}_{\tilde{n}}(-M_{cp}+1),\ldots, \right.  \tilde{f}_{\tilde{n}}(-M_{cp}+L-1)\right]^T, \tilde{n}\in \mathcal{U}_{2N}\right\},
%  \end{aligned}
  \label{Eqn:10}
\end{equation}
where $\mathcal{U}_{2N}=\{0,1,\ldots,2N\}$. In \eqref{Eqn:9}, the design of the basis signal $\tilde{f}_{\tilde{n}}(m)$ and the calculation of the coefficients $b_{i,n}\in\mathbf{b}_i=[b_{i,0},b_{i,1},\ldots,b_{i,N}]^T$ will be specified as follows.

To guarantee the smoothness of $\tilde{w}_i(m)$ and to obtain the \emph{N}-continuous signal, an example of designing $\tilde{f}_{\tilde{n}}(m)$ is given by
\begin{equation}
  \tilde{f}_{\tilde{n}}(m)=f^{(\tilde{n})}(m)g_h(m),
  \label{Eqn:11}
\end{equation}
for $m\in\mathcal{L}$, and $ \tilde{f}_{\tilde{n}}(m)=0$ for $m\in\mathcal{M}-\mathcal{L}$. In $\mathcal{L}$,
\begin{equation}
f^{(\tilde{n})}(m)= 1/M\left(j2\pi/M\right)^{\tilde{n}}\!\!\sum\limits_{k_r\in\mathcal{K}}\!\!\!{k^{\tilde{n}}_re^{j\varphi k_r}e^{j2\pi\frac{k_r}{M}m}},
  \label{Eqn:12}
\end{equation}
 and $g_h(m)$ is the right half part of a baseband-equivalent window function $g(m)$. Thus, the discontinuity at the adjacent point between two consecutive OFDM symbols can be eliminated by $\tilde{w}_i(m)$ from only a single side of the adjacent point. Under the constraint of $g_h(m)$, the interference caused by $\tilde{w}_i(m)$ can be limited to the front section of the OFDM symbol. In order to satisfy \eqref{Eqn:9}, $g(m)$ should be considered as a smooth and zero-edged window function, such as a triangular, Hanning, or Blackman window function. 

On the other hand, the linear combination coefficients $b_{i,n}$ in \eqref{Eqn:9} can be calculated as
\begin{equation}
  {\mathbf{b}}_i=\mathbf{P}^{-1}_{\tilde{f}}\begin{bmatrix}
     y_{i-1}(M)-y_i(-M_{cp})\\
     {\mathbf{P}}_1\mathbf{x}_{i-1}-\mathbf{P}_2\mathbf{x}_i
\end{bmatrix},
  \label{Eqn:13}
\end{equation}
where $\mathbf{P}_{\tilde{f}}$  is a $(N+1)\times(N+1)$ symmetric matrix related to ${\mathcal{Q}}$ with element  $\left\{\mathbf{P}_{\tilde{f}}\right\}_{n+1,v+1}$ is $\tilde{f}^{(n+v)}(-M_{cp})$, and element $\left\{{\mathbf{P}}_1\right\}_{n,r+1}=1/M\left(j2\pi k_r/M\right)^n$  and element  $\left\{\mathbf{P}_2\right\}_{n,r+1}=1/M\left(j2\pi k_r/M\right)^ne^{j\varphi k_r}$ for $n\neq 0$.

\begin{figure}[thbp]%[!t]
\centering
\includegraphics[width=3.6in]{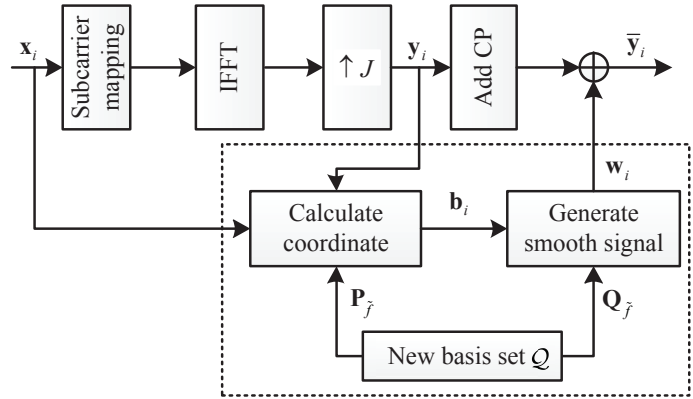}
\DeclareGraphicsExtensions.
\caption{Block diagram of the OFDM transmitter with the proposed low-interference TD-NC-OFDM.}
\label{Fig:3}
\end{figure}

Figure \ref{Fig:3} shows the block diagram of the proposed low-interference TD-NC-OFDM. According to $\mathcal{K}$ and $\mathcal{Q}$, the matrices $\mathbf{Q}_{\tilde{f}}=[\mathbf{q}_0, \mathbf{q}_1, \ldots, \mathbf{q}_N ]$, $\mathbf{P}^{-1}_{\tilde{f}}$,  $\mathbf{P}_{1}$, and  $\mathbf{P}_{2}$, can be calculated and stored in advance. Then, the data is first mapped and transformed to the time domain by the IFFT. Furthermore, the oversampled OFDM signal is appended by a CP. Finally, under the initialization of $\mathbf{y}_{-1}=\mathbf{0}$, the smooth signal is added onto the OFDM signal constructed by $M_s$ OFDM symbols $\mathbf{y}_i$ to generate the following transmit signal
\begin{equation}
  \bar{\mathbf{y}}_i=\left\{\begin{matrix}
      \mathbf{y}_i+\mathbf{Q}_{\tilde{f}}{\mathbf{b}}_i, & 0\leq i\leq M_s \\
      \mathbf{Q}_{\tilde{f}}{\mathbf{b}}_i, &  i=M_s+1
\end{matrix}\right..
  \label{Eqn:14}
\end{equation}

%=======================================================================================================================================================================
\section{Analysis of Spectrum and Complexity}
%-----------------------------------------------------------------------------------------------------------------------------------------------------------------------------------------------------------------------------------------------------------------------------------------------------
\subsection{Spectral Analysis}

According to the definition of PSD \cite{Ref8} and the relationship between spectral roll-off and continuity \cite{Ref16}, the PSD of the OFDM signal processed by the low-interference scheme is achieved as follows.

All the derivatives of the OFDM signal $s(t)$ are known to exist except for around the two edges and on the points between adjacent OFDM symbols. Meanwhile, except for these non-differentiable points, the smooth signal also possesses derivatives of all orders, according to the existence of all the derivatives of the basis function $\tilde{f}_n(m)$. Then, we assume that at non-differentiable points, the first \emph{N}-1 derivatives of the smoothed OFDM signal $\bar{s}(t)$ are continuous, and the \emph{N}th-order derivative $\bar{s}^{(N)}(t)$ has finite amplitude discontinuity. We also suppose that all the derivatives of $\bar{s}(t)$ approach zeroes , which corresponds to \eqref{Eqn:14}. Firstly, based on \cite{Ref16}, we can obtain
\begin{equation}
  \mathcal{F}\left\{\bar{s}(t)\right\}=\frac{1}{(j2\pi f)^{N-1}}\int\limits^{+\infty}_{-\infty}{\bar{s}^{(N-1)}(t)e^{-j2\pi ft}dt}. 
  \label{Eqn:15}
\end{equation}

Furthermore, since $\bar{s}^{(N)}(t)$ has finite amplitude discontinuities, by setting $u=\bar{s}^{(N-1)}(t)$ and $dv=e^{-j2\pi ft}dt$ in the above expression, we arrive at
\begin{align}
  \mathcal{F}\left\{\bar{s}(t)\right\}
  &=\frac{1}{(j2\pi f)^{N-1}}\left(\!\left.\frac{\bar{s}^{(N-1)}(t)e^{-j2\pi ft}}{-j2\pi f}\right|^{t=+\infty}_{t=-\infty}\!-\!\int\limits^{+\infty}_{-\infty}{\frac{s^{(N)}(t)e^{-j2\pi ft}}{-j2\pi f}dt}\!\right) \nonumber \\
  &=\frac{1}{(j2\pi f)^{N}}\int\limits^{+\infty}_{-\infty}{\bar{s}^{(N)}(t)e^{-j2\pi ft}dt}.
  \label{Eqn:16}
  \end{align}

Because $\bar{s}^{(N)}(t)$ has finite amplitude discontinuities at the adjacent points, from \eqref{Eqn:2}, $\bar{s}^{(N)}(t)$ can be written as
\begin{equation}
  \bar{s}^{(N)}(t)=\sum\limits^{+\infty}_{i=-\infty}{\bar{y}^{(N)}_i\left(t-iT\right)}.
  \label{Eqn:17}
\end{equation}

It is inferred in \eqref{Eqn:17} that the \emph{N}th derivative $\bar{y}^{(N)}_i(t)$ can be assumed being windowed by the rectangular function $R(t)$. Therefore, based on the definition of PSD, \eqref{Eqn:16}, and \eqref{Eqn:17}, the PSD of $\bar{s}(t)$ can be expressed as
 \begin{align}
 {\Psi}(f) \!=\!\! \lim\limits_{U\rightarrow \infty}{\frac{1}{2UT}E\left\{\left|\mathcal{F}\left\{
  \sum\limits^{U-1}_{i=-U}{\!\frac{\bar{y}^{(N)}_{i}(t-iT)}{\left(j2\pi f\right)^N}}\right\}\right|^2\!\right\}}  % \nonumber \\
  \!=\!\! \lim\limits_{U\rightarrow \infty}{\!\frac{1}{2UT}E\left\{\left|\sum\limits^{U-1}_{i=-U}{\frac{\mathcal{F}
   \left\{\bar{y}^{(N)}_{i}(t)\right\}}{\left(j2\pi f\right)^N}e^{-j2\pi fiT}}\right|^2\!\right\}}.
  \label{Eqn:18}
\end{align}

Eq. \eqref{Eqn:18} indicates that the spectrum of the TD-NC-OFDM signal is related to the expectation of $\bar{y}^{(N)}_i(t)$ and $f^{-N}$. In this paper, the conventional Blackman window function is used as an example, given as    $g(t)=0.42-0.5\cos{(2\pi \rho t)}+0.08\cos{(4\pi \rho t)}$ where $\rho=1/\left((2L-2)T_{samp}\right)$. By substituting \eqref{Eqn:9}, \eqref{Eqn:11}, \eqref{Eqn:12}, and \eqref{Eqn:14} into \eqref{Eqn:18}, the PSD of the smoothed OFDM signal in low-interference scheme is expressed by
  \begin{align}
  {\Psi}(f)
         &=\lim\limits_{i\rightarrow \infty}\frac{1}{2UT}E\Bigg\{\bigg|\sum\limits^{U-1}_{i=-U}e^{-j2\pi fiT}\left(j2\pi f\right)^{-N} 
          \sum\limits_{k_r\in \mathcal{K}}{\left(\frac{j2\pi k_r}{T_s}\right)^Nx_{i,k_r}\mathrm{sinc}\left(f_r(1+\beta)\right)e^{j\pi f_r(1-\beta)}}  \nonumber \\
        & \quad +\frac{1}{T}\sum\limits^{N}_{n=0}{b}_{i,n}\sum\limits^{N}_{\bar{n}=0}\left(\begin{matrix}
            N \\ \bar{n} \end{matrix}\right)\sum\limits_{k_r\in \mathcal{K}}\left(\frac{j2\pi k_r}{T_s}\right)^{N-\bar{n}+n} 
           \int\limits^{-T_{cp}+T_p}_{-T_{cp}}{g^{(\bar{n})}_{h}(t)e^{j2\pi f_r t/T_s}dt}\bigg|^2\Bigg\} \nonumber \\
        &=\lim\limits_{i\rightarrow \infty}\frac{1}{2UT}E\Bigg\{\bigg|\sum\limits^{U-1}_{i=-U}e^{-j2\pi fiT}\left(fT_s\right)^{-N} 
         \times \sum\limits_{k_r\in \mathcal{K}}{k^N_rx_{i,k_r}\mathrm{sinc}\left(f_r(1+\beta)\right)e^{j\pi f_r(1-\beta)}}   \nonumber \\
        & \quad +\!\frac{1}{T}\!\sum\limits^{N}_{n=0}{b}_{i,n}\!\sum\limits^{N}_{\bar{n}=0}\!\!\left(\begin{matrix}
            N \\ \bar{n} \end{matrix}\right)\!\!\left(\!\dfrac{j2\pi}{T_s}\!\right)^{n-\bar{n}} %\\
         \!\!\! \sum\limits_{k_r\in \mathcal{K}}\!{k^{N-\bar{n}+n}_rG_{\bar{n}}(f)}\bigg|^2\!\Bigg\}
  \label{Eqn:19}
  \end{align}
where $G_{\bar{n}}(f)$ is given by
 \begin{eqnarray*}
  \!\!\!&\!\!\! G_{\bar{n}}(f)=e^{j\tilde{f}_r}\Bigg(0.42^{(\bar{n})}T_p\mathrm{sinc}\left(\!\mu f_r\right)-\dfrac{0.5(2\pi\rho)^{\bar{n}}\cos{\left(\pi\mu f_r\right)}}{1-\left(\rho T_s/f_r\right)^2} \left(\dfrac{\cos{\left(\pi\bar{n}/2\right)}}{j\pi f_r/T_s}-\pi\rho\dfrac{\sin{\left(\pi\bar{n}/2\right)}}{\left(\pi f_r/T_s\right)^2}\right) \\
   \!\!\!&\!\!\! + \dfrac{0.08(4\pi\rho)^{\bar{n}}\sin{\left(\pi\mu f_r\right)}}{1-\left(2\rho T_s/f_r\right)^2}
    \!\left(\!\!\dfrac{\cos{\left(\pi\bar{n}/2\right)}}{\pi f_r/T_s}\!-j2\pi\rho\dfrac{\sin{\left(\pi\bar{n}/2\right)}}{\left(\pi f_r/T_s\right)^2}\!\right)\!\!\!\Bigg),
\end{eqnarray*}
where $\text{sinc}(x)\triangleq \sin(\pi x)/(\pi x)$, and $\tilde{f}_r=\pi f_r(T_p-2T_{cp})/T_s$ with $f_r=k_r-T_sf$, $T_p=(L-1)T_{samp}$, and $\mu=T_p/T_s$. 

\begin{figure}[thbp]%[!t]
\centering
\includegraphics[width=4in]{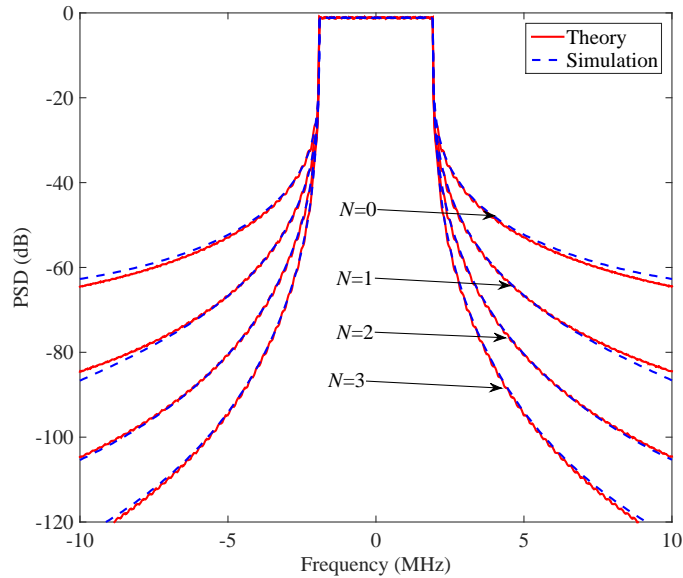}
\DeclareGraphicsExtensions.
\caption{PSD comparison between the analytical and simulation results for the TD-NC-OFDM signal in the low-interference TD-NC-OFDM with \emph{L}=144.}
\label{Fig:4}
\end{figure}

Eq. \eqref{Eqn:19} shows that the power spectral roll-off of the smoothed signal, whose first \emph{N}-1 derivatives are continuous, decays with $f^{-2N-2}$. Moreover, $G_{\bar{n}}(f)$ reveals that the sidelobe is affected by the length of $g(t)$, so that a rapid sidelobe decaying can be achieved by increasing the length of $g(t)$. Figure \ref{Fig:4} compares the theoretical and simulation results of the low-interference scheme with a window length of 144. It is shown that the simulation results match well with the theoretical analyses.

%-----------------------------------------------------------------------------------------------------------------------------------------------------------------------------------------------------------------------------------------------------------------------------------------------------
\subsection{Complexity Comparison}

Firstly, we consider the complexity of the transmitter. In NC-OFDM, its frequency-domain precoder requires $2K^2$ complex multiplications and $2K^2$ complex additions as indicated in \eqref{Eqn:6}. In TD-NC-OFDM \cite{Ref15}, $2NK+(N+1)(2N+1)$ complex multiplications and $2NK+N(2N+1)$ complex additions are required. However, for the generation and overlapping of the smooth signal, $M(N+1)$ complex multiplications and $M(N+1)$ complex additions are needed. By shortening the length of the smooth signal, the low-interference scheme just requires $L(N+1)$ complex multiplications and $L(N+1)$ complex additions. At the same time, the complexity of calculating its linear combination coefficients in \eqref{Eqn:13} is $2NK+(N+1)^2$ complex multiplications and $2NK+N^2+1$ complex additions.

Secondly, the complexity of the receiver is shown in Table \ref{Tab1}. For NC-OFDM, an iterative signal recovery algorithm \cite{Ref9} is used to eliminate the interference. Due to the equivalence between NC-OFDM and TD-NC-OFDM, the signal recovery algorithm is also desired in TD-NC-OFDM with identical complexity. In the low-interference scheme, the receiver is the same as in original OFDM without extra signal recover processing.

Assume that a complex addition is equivalent to two real additions; a complex multiplication to four real multiplications plus two real additions; and a real-complex multiplication to two real multiplications. The complexity comparison among NC-OFDM, TD-NC-OFDM and the low-interference scheme is shown in Table \ref{Tab1}, where $L_R$ denotes the number of iterations in the signal recovery algorithm \cite{Ref9}.

\begin{table}[!ht]
\footnotesize
\begin{center}
\caption{Complexity Comparison among NC-OFDM, TD-NC-OFDM and its Low-Interference Scheme}
\label{Tab1}
\begin{tabular*}{13.65cm}{|m{3.5cm}|m{1.5cm}|m{3.45cm}|m{3.45cm}|}
\hline 
  \multicolumn{2}{|c|}{\bfseries Scheme} & \bfseries{Real multiplication} & \bfseries{Real addition} \\
\hline
  \bfseries NC-OFDM & \bfseries{Transmitter} & $O(8K^2)$ & $O(8K^2)$  \\
  {} & \bfseries{Receiver} & $O(16(N+1)K L_R)$ & $O(16(N+1)K L_R)$ \\
\hline
\bfseries TD-NC-OFDM & \bfseries{Transmitter} & $O(8NK+4(N+1)M)$ & $O(8NK+4(N+1)M)$ \\
  {} & \bfseries{Receiver} &  $O(16(N+1)K L_R)$ & $O(16(N+1)K L_R)$ \\
\hline
\bfseries  Low-interference scheme & \bfseries{Transmitter} & $O(8NK+4(N+1)L)$ & $O(8NK+4(N+1)L)$ \\
  {} & \bfseries{Receiver} & $0$  &  $0$  \\
\hline
  \multicolumn{2}{|c|}{\bfseries \emph{M}-point IFFT/FFT} & $O(2M\log_2 M)$ & $O(2M\log_2 M)$ \\
\hline
\end{tabular*}
\end{center}
\end{table}

The low-interference scheme considerably suppresses the sidelobes shown in Figure \ref{Fig:4}. Simultaneously, since the length of the smooth signal \emph{L} is often as short as the CP length or shorter in real systems, the low-interference scheme is of lower transmitter complexity than NC-OFDM and the conventional TD-NC-OFDM. Moreover, its complexity is comparable to \emph{M}-point IFFT. On the other hand, the low-interference scheme just requires the receiver of original OFDM, and avoiding extra processing in NC-OFDM and TD-NC-OFDM receivers.

%-----------------------------------------------------------------------------------------------------------------------------------------------------------------------------------------------------------------------------------------------------------------------------------------------------
\subsection{SINR Analysis}

One disadvantage of NC-OFDM is that the transmit signal is easily interfered by the smooth signal. Thus, a measure is needed to evaluate the interference in terms of the SNR loss. In this section, we investigate the SINR of NC-OFDM, and demonstrate the effectiveness of the proposed low-interference scheme in reducing the SNR loss, based on the analysis of the average power of the smooth signal.

In a multipath channel with time-domain coefficients $h_{\tilde{l}}$ in the $\tilde{l}$th path, the \emph{i}th received time-domain OFDM symbol $r_i(t)$ is given by
\begin{equation}
  r_i(t)=\sum\limits^{\tilde{L}}_{\tilde{l}=1}{h_{\tilde{l}}\bar{y}_i(t-\tau_{\tilde{l}})+n_i(t)}
  \label{Eqn:20}
\end{equation}
where $\tau_{\tilde{l}}$ is the time delay in the $\tilde{l}$th path, and $n_i(t)$ is the AWGN noise with mean zero and variance $\sigma^2_n$. 

Because $\mathbf{x}_i$ is uncorrelated, i.e., $E\left\{\mathbf{x}_i\mathbf{x}^H_i\right\}=\mathbf{I}_K$ and $E\left\{\mathbf{x}_{i-1}\mathbf{x}^H_i\right\}=\mathbf{0}_{K\times K}$, we can obtain $E\left\{\mathbf{x}^H_i\mathbf{x}_i\right\}=
\mathrm{Tr}\left\{E\left\{\mathbf{x}_i\mathbf{x}^H_i\right\}\right\}=K$. Thus, the average power of the OFDM symbol vector $\mathbf{y}_i$ is
\begin{equation}
  E\left\{\mathbf{y}^H_i\mathbf{y}_i\right\}
  =\frac{1}{M^2}E\left\{\mathbf{x}^H_i\mathbf{F}_f\mathbf{F}^H_f\mathbf{x}_i\right\}
  =E\left\{\mathbf{x}^H_i\mathbf{x}_i\right\}
  ={K}/{M}.
  \label{Eqn:21}
\end{equation}

\begin{figure}[thbp]%[!t]
\centering
\includegraphics[width=4in]{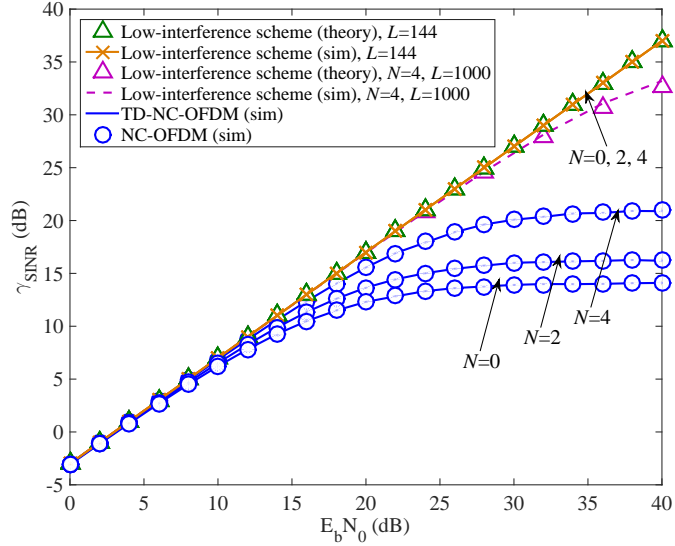}
\DeclareGraphicsExtensions.
\caption{A time-domain illustration of the effect of the smooth signal on the multipath channel without power attenuation and Gaussian noise.}
\label{Fig:5}
\end{figure}

To mitigate the performance degradation in conventional TD-NC-OFDM, the proposed low-interference scheme is analyzed by exploring the distribution of $\tilde{w}_i(l)$ in the multipath fading channel. As illustrated in Figure \ref{Fig:5}, different channel paths with varying time delays lead to varying $\tilde{w}_i(l)$. With the increased time delay, the delayed tail of $\tilde{w}_i(l)$ is prolonged and the interference increases. The interferences $\tilde{w}_i(l)$ are composed of the delayed tails in all the paths, whose powers are much smaller than the conventional TD-NC-OFDM and NC-OFDM.

Thus, the average power of $\tilde{w}_i(l)$ in the $\tilde{l}$th path is calculated by
\begin{align}
 E\left\{\left(h_{\tilde{l}}\tilde{\mathbf{w}}_i\right)^H h_{\tilde{l}}\tilde{\mathbf{w}}_i\right\} 
 &= \mathrm{Tr}\left\{E\left\{h_{\tilde{l}}\tilde{\mathbf{w}}^H_i
                               \tilde{\mathbf{w}}_ih^H_{\tilde{l}}\right\}\right\} 
 = E\left\{\left|h_{\tilde{l}}\right|^2\right\}
           \mathrm{Tr}\left\{E\left\{\tilde{\mathbf{w}}^H_i\tilde{\mathbf{w}}_i\right\}\right\} \nonumber\\
 &= \frac{2}{M^2}E\left\{\left|h_{\tilde{l}}\right|^2\right\}
     \mathrm{Tr}\left\{\mathbf{P}^{-1}_{\tilde{f}}\mathbf{B}_2\mathbf{B}^H_2
     \left(\mathbf{P}^{-1}_{\tilde{f}}\right)^H
     \mathbf{Q}^H_{\tilde{f}_{\tilde{l}}}\mathbf{Q}_{\tilde{f}_{\tilde{l}}}\right\}
  \label{Eqn:22}
\end{align}
where $\mathbf{g}_{\tilde{l}}$ is the delayed tail of $\mathbf{g}$ in the $\tilde{l}$th path and $\mathbf{Q}_{\tilde{f}_{\tilde{l}}}=\mathbf{g}_{\tilde{l}}\mathbf{Q}_f$.

The derivation of $\mathrm{Tr}\left\{E\left\{\tilde{\mathbf{w}}^H_i\tilde{\mathbf{w}}_i\right\}\right\}$ is shown in Appendix A.

Therefore, in all the paths, the average power of these delayed tails is expressed by
\begin{equation}
  \sum\limits^{\tilde{L}}_{\tilde{l}=1}
  {E\left\{\left(h_{\tilde{l}}\tilde{\mathbf{w}}_i\right)^Hh_{\tilde{l}}\tilde{\mathbf{w}}_i\right\}} 
 = \frac{2}{M^2}\sum\limits^{\tilde{L}}_{\tilde{l}=1}{E\left\{\left|h_{\tilde{l}}\right|^2\right\}
     \mathrm{Tr}\left\{\mathbf{P}^{-1}_{\tilde{f}}\mathbf{B}_2\mathbf{B}^H_2
     \left(\mathbf{P}^{-1}_{\tilde{f}}\right)^H
     \mathbf{Q}^H_{\tilde{f}_{\tilde{l}}}\mathbf{Q}_{\tilde{f}_{\tilde{l}}}\right\}}.
  \label{Eqn:23}
\end{equation}

Finally, from \eqref{Eqn:21}-\eqref{Eqn:23}, the received SINR ${\gamma}_{SINR}$ is obtained by
  \begin{align}
    {\gamma}_{SINR} 
    &=\dfrac{\sum\limits^{\tilde{L}}_{\tilde{l}=1}
  {E\left\{\left|h_{\tilde{l}}\right|^2\right\}E\left\{\mathbf{y}^H_i\mathbf{y}_i\right\}}}
  {\sigma^2_n+\sum\limits^{\tilde{L}}_{\tilde{l}=1}
  {E\left\{\left(h_{\tilde{l}}\tilde{\mathbf{w}}_i\right)^Hh_{\tilde{l}}\tilde{\mathbf{w}}_i\right\}}}  \nonumber \\
%  & =\dfrac{K/M\sum\limits^{\tilde{L}}_{\tilde{l}=1}{E\left\{\left|h_{\tilde{l}}\right|^2\right\}}}
%                           {\sigma^2_n
%                           +\dfrac{2}{M^2}\sum\limits^{\tilde{L}}_{\tilde{l}=1}{E\left\{\left|h_{\tilde{l}}\right|^2\right\}
%                           \mathrm{Tr}\left\{\mathbf{P}^{-1}_{\tilde{f}}\mathbf{B}_2\mathbf{B}^H_2
%                           \left(\mathbf{P}^{-1}_{\tilde{f}}\right)^H
%                           \mathbf{Q}^H_{\tilde{f}_{\tilde{l}}}
%                           \mathbf{Q}_{\tilde{f}_{\tilde{l}}}\right\}}}  \nonumber \\
  & =\dfrac{K/M}{\dfrac{\sigma^2_n}
                           {\sum\limits^{\tilde{L}}_{\tilde{l}=1}{E\left\{\left|h_{\tilde{l}}\right|^2\right\}}}
                           \!+\!\dfrac{2\sum\limits^{\tilde{L}}_{\tilde{l}=1}{E\left\{\left|h_{\tilde{l}}\right|^2\right\}
                           \mathrm{Tr}\!\!\left\{\mathbf{P}^{-1}_{\tilde{f}}\mathbf{B}_2\mathbf{B}^H_2
                           \left(\mathbf{P}^{-1}_{\tilde{f}}\right)^H
                           \!\!\mathbf{Q}^H_{\tilde{f}_{\tilde{l}}}
                           \mathbf{Q}_{\tilde{f}_{\tilde{l}}}\!\!\right\}}}
                           {M^2\sum\limits^{\tilde{L}}_{\tilde{l}=1}{E\left\{\left|h_{\tilde{l}}\right|^2\right\}}}}.
   \label{Eqn:24}
 \end{align}

\begin{figure}[h]%[!t]
\centering
\includegraphics[width=4in]{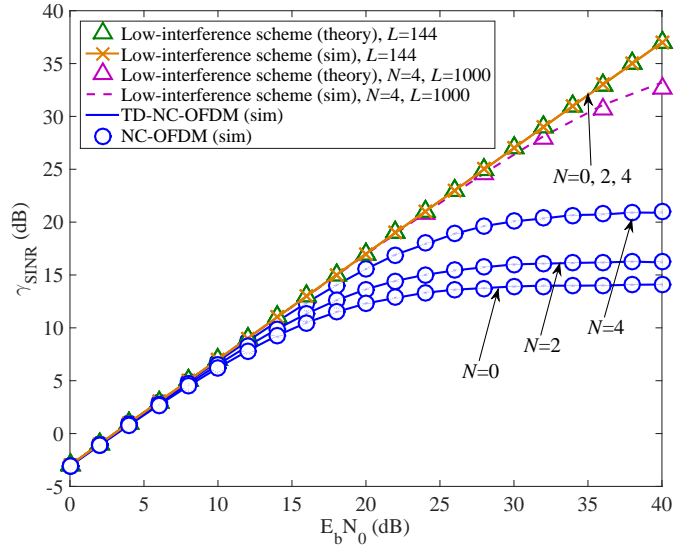}
\DeclareGraphicsExtensions.
\caption{SINR analysis and simulations of the TD-NC-OFDM signals in the EVA fading channel, where the signal is modulated by 16-QAM.}
\label{Fig:6}
\end{figure}

Figure \ref{Fig:6} compares theoretical analysis in \eqref{Eqn:24} and the simulation results in terms of the SINR. The simulated Rayleigh channel employs the Extended Vehicular A (EVA) channel model \cite{Ref17}, whose excess tap delay is [0, 30, 150, 310, 370, 710, 1090, 1730, 2510] ns with relative power [0, -1.5, -1.4, -3.6, -0.6, -9.1, -7, -12, -16.9] dB. It is shown that the theoretical analysis aligns well with the simulations. It also reveals that the SNR loss is negligible for the low-interference scheme. Moreover, the low-interference scheme has much better SINR than NC-OFDM and TD-NC-OFDM. In general, even if the length of $\tilde{w}_i(l)$ is increased, there is no extra need for signal recovery, which reduces the heavy computation load in original NC-OFDM.

%=======================================================================================================================================================================
\section{Numerical Results}

This section presents simulation results to evaluate the PSD, complexity, and BER performance of NC-OFDM, TD-NC-OFDM and proposed low-interference schemes. Simulations are performed in a baseband-equivalent OFDM system with 256 subcarriers mapped onto the subcarrier index set $\left\{-128,-127,\ldots,127\right\}$. 16-QAM digital modulation is employed with a symbol period $T_s=1/15$ms, time-domain oversampling interval $T_{samp}=T_s/2048$ and CP duration $T_{cp}=144T_{samp}$. The PSD is evaluated by Welch's averaged periodogram method with a 2048-sample Hanning window and 512-sample overlap after observing 105 symbols. To investigate the BER performance, the signal is transmitted through the Extended Vehicular A (EVA) channel model.

\begin{figure}[htbp]%[!t]
\centering
\includegraphics[width=4in]{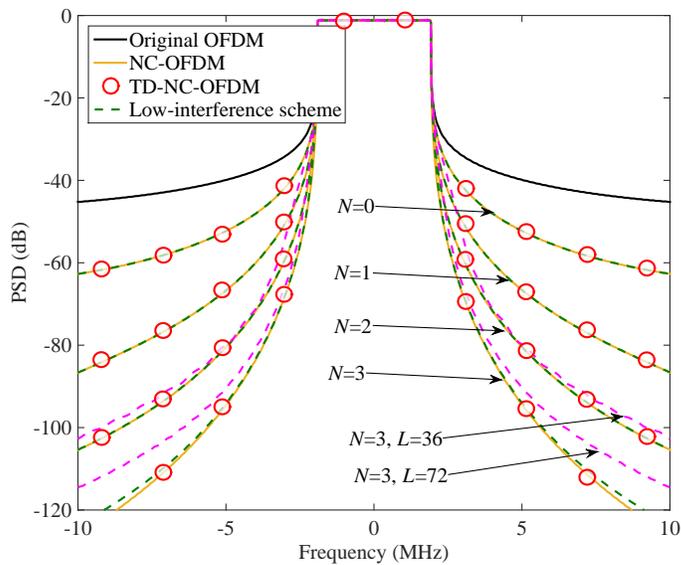}
\DeclareGraphicsExtensions.
\caption{PSDs of the transmit signals of NC-OFDM, TD-NC-OFDM and its low-interference scheme with varying \emph{N} and varying \emph{L}.}
\label{Fig:7}
\end{figure}

Figure \ref{Fig:7} compares the PSD of NC-OFDM transmit signals with different \emph{N} and different \emph{L}. As \emph{N} increases, the sidelobe suppression performance is further improved in the three methods. Moreover, the low-interference scheme can obtain as good sidelobe suppression performance as the conventional TD-NC-OFDM and NC-OFDM. Figure \ref{Fig:7} also shows that with the increase of \emph{L}, a steeper spectral roll-off can be obtained in the low-interference scheme. With a relatively small \emph{L}, the sidelobe suppression of the low-interference scheme can approach that of TD-NC-OFDM, such as \emph{N}=2 and 3 with \emph{L}=144.

\begin{figure}[htbp]%[!t]
\centering
\includegraphics[width=4in]{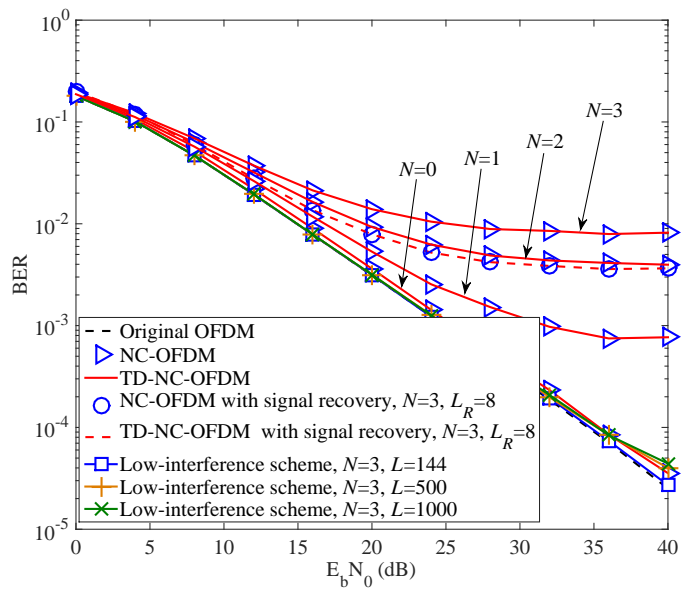}
\DeclareGraphicsExtensions.
\caption{BERs of NC-OFDM, TD-NC-OFDM and its low-interference scheme with varying \emph{N} and varying \emph{L} in the EVA channel.}
\label{Fig:8}
\end{figure}

Figure \ref{Fig:8} presents the BER performances of NC-OFDM, TD-NC-OFDM and its low-interference scheme with varying \emph{N} and varying \emph{L} in the EVA channel. It is shown that the BER performance of the received signal is significantly degraded as \emph{N} increases in NC-OFDM and TD-NC-OFDM. By contrast, the low-interference scheme causes slight BER performance degradation. Compared to the BER performance of NC-OFDM and TD-NC-OFDM with the high-complexity signal recovery \cite{Ref9}, the increased length of the smooth signal just results in slight performance degradation for the low SNR loss in the low-interference scheme as mentioned in Section 4.3. Meanwhile, the low-interference scheme exhibits promising sidelobe suppression performance as shown in Figure \ref{Fig:7}.  

%\begin{figure}[!ht]
%\centering
%\begin{minipage}[c]{0.48\textwidth}
%\centering
%\includegraphics[width=3.3in]{Figure7.eps}
%\end{minipage}
%\hspace{0.02\textwidth}
%\begin{minipage}[c]{0.48\textwidth}
%\centering
%\includegraphics[width=3.3in]{Figure8.eps}
%\end{minipage}\\[3mm]
%\begin{minipage}[t]{0.48\textwidth}
%\centering
%\caption{PSDs of the transmit signals of NC-OFDM, TD-NC-OFDM and its low-interference scheme with varying \emph{N} and varying \emph{L}.}
%\label{Fig:7}
%\end{minipage}
%\hspace{0.02\textwidth}
%\begin{minipage}[t]{0.48\textwidth}
%\centering
%\caption{BERs of NC-OFDM, TD-NC-OFDM and its low-interference scheme with varying \emph{N} and varying \emph{L} in the EVA channel.}
%\label{Fig:8}
%\end{minipage}
%\end{figure}

%=======================================================================================================================================================================
\section{Conclusion}

In this paper, a low-interference TD-NC-OFDM was proposed to reduce the interference and implementation complexity as opposed to the original TD-NC-OFDM and NC-OFDM. By adding the smooth signal, the \emph{N}-continuous signal was obtained by the low-interference scheme. The smooth signal was designed by the linear combination of a basis set, which is generated by rectangularly pulsed OFDM basis signals truncated by a smooth window. Furthermore, using the continuity criterion, a closed-form spectrum expression was derived in the low-interference TD-NC-OFDM, which had more rapid decaying than \cite{Ref16}. Then the complexity of the low-interference scheme was measured. The received SINR is also measured by deriving the closed-form expression. Simulation results showed that the low-interference scheme was capable of effectively suppressing sidelobes as well as NC-OFDM and TD-NC-OFDM but with much better BER performance and much lower complexity. In this sense, the low-interference TD-NC-OFDM is a promising alternative to conventional NC-OFDM in future cognitive radio and carrier aggregation combined 5G systems.

\section*{Acknowledgment}
This work was supported by the open research fund of National Mobile Communications Research Laboratory, Southeast University (No. 2013D05).

%%%%%%%%%%%%%%%%%%%%%%%%%%%%%%%%%%%%%%%%%%%%%%%%%%%%%%%
%%% Appendix sections. 附录章节, 非必选
%%%%%%%%%%%%%%%%%%%%%%%%%%%%%%%%%%%%%%%%%%%%%%%%%%%%%%%
\section*{Appendix: Derivation of $\mathrm{Tr}\left\{E\left\{\tilde{\mathbf{w}}^H_i\tilde{\mathbf{w}}_i\right\}\right\}$}

For ease of exposition, Eq. \eqref{Eqn:13} is rewritten as
\begin{equation}
  \mathbf{b}_i
  =\mathbf{P}^{-1}_{\tilde{f}}\left({\mathbf{P}}_1\mathbf{x}_{i-1}-\mathbf{P}_2\mathbf{x}_i\right)
   \label{Eqn:25}
\end{equation}
where ${\mathbf{P}}_1$ and $\mathbf{P}_2$ include a row of $n=0$. 

According to the construction of ${\mathbf{P}}_1$ and $\mathbf{P}_{2}$, $\mathrm{Tr}\left\{E\left\{\tilde{\mathbf{w}}^H_i\tilde{\mathbf{w}}_i\right\}\right\}$ can be expressed as
\begin{eqnarray}
 \mathrm{Tr}\left\{E\left\{\tilde{\mathbf{w}}^H_i\tilde{\mathbf{w}}_i\right\}\right\} %\nonumber\\
  &=& \mathrm{Tr}\left\{E\left\{\mathbf{Q}_{\tilde{f}}\mathbf{b}_i
                         \mathbf{b}^H_i\mathbf{Q}^H_{\tilde{f}}\right\}\right\} \nonumber \\
  &=& \mathrm{Tr}\left\{E\left\{\mathbf{Q}_{\tilde{f}}\mathbf{P}^{-1}_{\tilde{f}}{\mathbf{P}}_{1}
                         \mathbf{x}_{i-1}\mathbf{x}^H_{i-1}{\mathbf{P}}^H_{1}
                         \left(\mathbf{P}^{-1}_{\tilde{f}}\right)^H\mathbf{Q}^H_{\tilde{f}}\right\}\right\} \nonumber\\
   & &  + \mathrm{Tr}\left\{E\left\{\mathbf{Q}_{\tilde{f}}\mathbf{P}^{-1}_{\tilde{f}}\mathbf{P}_{2}
                         \mathbf{x}_{i}\mathbf{x}^H_{i}\mathbf{P}^H_{2}
                         \left(\mathbf{P}^{-1}_{\tilde{f}}\right)^H\mathbf{Q}^H_{\tilde{f}}\right\}\right\} \nonumber \\
  &=& \mathrm{Tr}\left\{\mathbf{P}^{-1}_{\tilde{f}}{\mathbf{P}}_{1}{\mathbf{P}}^H_{1}
                         \left(\mathbf{P}^{-1}_{\tilde{f}}\right)^H
                         \mathbf{Q}^H_{\tilde{f}}\mathbf{Q}_{\tilde{f}}\right\}% \nonumber\\
     + \mathrm{Tr}\left\{\mathbf{P}^{-1}_{\tilde{f}}\mathbf{P}_{2}\mathbf{P}^H_{2}
                         \left(\mathbf{P}^{-1}_{\tilde{f}}\right)^H
                         \mathbf{Q}^H_{\tilde{f}}\mathbf{Q}_{\tilde{f}}\right\}.
   \label{Eqn:26}
\end{eqnarray}

Then, we rewrite ${\mathbf{P}}_1$ and $\mathbf{P}_2$ as $\mathbf{P}_1={1}/{M}\mathbf{B}_2$ and $\mathbf{P}_2={1}/{M}\mathbf{B}_2\mathbf{\Phi}$ with $\left\{\mathbf{B}_2\right\}_{n+1,r+1}=\left(j 2\pi k_r/M\right)^n$. We obtain 
\begin{equation}
\mathbf{B}_1=\mathbf{\Phi}^H\mathbf{B}^T_2,
   \label{Eqn:27}
\end{equation}
with $\left\{\mathbf{B}_1\right\}_{r+1,n+1}=\left(j2\pi k_r/M\right)^ne^{-j\varphi k_r}$.

According to \eqref{Eqn:27}, we arrive at 
\begin{eqnarray}
 {\mathrm{Tr}\left\{E\left\{\tilde{\mathbf{w}}^H_i\tilde{\mathbf{w}}_i\right\}\right\}}% \nonumber \\
  &=& \frac{1}{M^2}\mathrm{Tr}\left\{\mathbf{P}^{-1}_{\tilde{f}}\mathbf{B}_2\mathbf{B}^H_2
                                      \left(\mathbf{P}^{-1}_{\tilde{f}}\right)^H
                                      \mathbf{Q}^H_{\tilde{f}}\mathbf{Q}_{\tilde{f}}\right\} %\nonumber \\
     + \frac{1}{M^2}\mathrm{Tr}\left\{\mathbf{P}^{-1}_{\tilde{f}}\mathbf{B}_2\mathbf{\Phi}\mathbf{\Phi}^H
                                      \mathbf{B}^H_2\left(\mathbf{P}^{-1}_{\tilde{f}}\right)^H
                                      \mathbf{Q}^H_{\tilde{f}}\mathbf{Q}_{\tilde{f}}\right\} \nonumber \\
  &=& \frac{2}{M^2}\mathrm{Tr}\left\{\mathbf{P}^{-1}_{\tilde{f}}\mathbf{B}_2\mathbf{B}^H_2
                                      \left(\mathbf{P}^{-1}_{\tilde{f}}\right)^H
                                      \mathbf{Q}^H_{\tilde{f}}\mathbf{Q}_{\tilde{f}}\right\}.
   \label{Eqn:28}
\end{eqnarray}



%%%%%%%%%%%%%%%%%%%%%%%%%%%%%%%%%%%%%%%%%%%%%%%%%%%%%%%
%%% Supplements. 补充材料, 非必选
%%%%%%%%%%%%%%%%%%%%%%%%%%%%%%%%%%%%%%%%%%%%%%%%%%%%%%%


\begin{thebibliography}{99}

\bibitem{Ref1}
Hwang T, Yang C, Wu G, et al. OFDM and its wireless applications: A survey. IEEE Trans Veh Technol, 2009, 58: 1673-1694
\bibitem{Ref18}
Wang C X, Haider F, Gao X, et al. Cellular architecture and key technologies for 5G wireless communication networks. IEEE Commun Mag, 2014, 52: 122-130
\bibitem{Ref19}
Wang Y, Li J, Huang L, et al. 5G mobile: spectrum broadening to higher-frequency bands to support high data rates. IEEE Trans Veh Technol, 2014, 9: 39-46
\bibitem{Ref20}
Hong X, Wang J, Wang C X, et al. Cognitive radio in 5G: a perspective on energy-spectral efficiency trade-off. IEEE Commun Mag, 2014, 52: 46-53
\bibitem{Ref21}
Yuan G, Zhang X, Wang W, et al. Carrier aggregation for LTE-advanced mobile communication systems. IEEE Commun Mag, 2010, 48: 88-93
\bibitem{Ref2}
Bogucka H, Wyglinski A M, Pagadarai S, et al. Spectrally agile multicarrier waveforms for opportunistic wireless access. IEEE Commun Mag, 2011, 49: 108-115
\bibitem{Ref3}
Weiss T, Hillenbrand J, Krohn A, et al. Mutual interference in OFDM-based spectrum pooling systems. In: The 59th IEEE Vehicular Technology Conference Spring (VTC 2004-Spring), Milan, 2004. 1873-1877
\bibitem{Ref4}
Brandes S, Cosovic I, Schnell M. Reduction of out-of-band radiation in OFDM systems by insertion of cancellation carriers. IEEE Commun Lett, 2006, 10: 420-422
\bibitem{Ref5}
Qu D, Wang Z, Jiang T. Extended active interference cancellation for sidelobe suppression in cognitive radio OFDM systems with cyclic prefix. IEEE Trans Veh Technol, 2010, 59: 1689-1695
\bibitem{Ref6}
Ma M, Huang X, Jiao B, et al. Optimal orthogonal precoding for power leakage suppression in DFT-based systems. IEEE Trans Commun, 2011, 59: 844-853
\bibitem{Ref7}
Zhang J, Huang X, Cantoni A, et al. Sidelobe suppression with orthogonal projection for multicarrier systems. IEEE Trans Commun, 2012, 60: 589-599
\bibitem{Ref8}
Chung C D. Spectrally precoded OFDM. IEEE Trans Commun, 2006, 54: 2173-2185
\bibitem{Ref9}
Beek de van J, Berggren F. \emph{N}-continuous OFDM. IEEE Commun Lett, 2009, 13: 1-3
\bibitem{Ref10}
Beek de van J, Berggren F. EVM-constrained OFDM precoding for reduction of out-of-band emission. In: The 70th Vehicular Technology Conference Fall (VTC 2009-Fall), Anchorage, 2009. 1-5
\bibitem{Ref11}
Beek de van J. Sculpting the multicarrier spectrum: a novel projection precoder. IEEE Commun Lett. 2009, 13: 881-883
\bibitem{Ref12}
Ohta M, Iwase A, Yamashita K. Improvement of the error characteristics of an \emph{N}-continuous OFDM system with low data channels by SLM. In: The 2011 IEEE International Conference on Communications (ICC 2011), Kyoto, 2011. 1-5
\bibitem{Ref13}
Ohta M, Okuno M, Yamashita K. Receiver iteration reduction of an \emph{N}-continuous OFDM system with cancellation tones. In: The 2011 IEEE Global Telecommunications Conference (GLOBECOM 2011), Kathmandu, 2011. 1-5
\bibitem{Ref14}
Zheng Y, Zhong J, Zhao M, et al. A precoding scheme for \emph{N}-continuous OFDM. IEEE Commun Lett, 2012, 16: 1937-1940
\bibitem{Ref15}
Wei P, Dan L, Xiao Y, et al. A Low-Complexity Time-Domain Signal Processing Algorithm for \emph{N}-continuous OFDM. In: The 2013 IEEE International Conference on Communications. (ICC 2013), Budapest, 2013. 5754-5758
\bibitem{Ref16}
Bracewell R, The Fourier Transform and its applications, 2nd ed. New York: McGraw-Hill, 1978. 143-146
\bibitem{Ref17}
User Equipment (UE) radio transmission and reception (Release 12), 3GPP TS 36.101, v12.3.0, 2014. [Online]. Available: http: //www.3gpp.org/

\end{thebibliography}
\end{document}